\documentclass[
 reprint,
 amsmath,amssymb,
 aps, superscriptaddress, pre
]{revtex4-1}
\usepackage{graphicx}
\usepackage{amsmath}
\usepackage{amssymb}
\usepackage{enumerate}
\usepackage{bm}
\usepackage[justification=raggedright,singlelinecheck=false]{caption}
\usepackage[justification=raggedright,singlelinecheck=false]{subcaption}
\usepackage{cleveref}
\usepackage{verbatim}
\usepackage{makecell}
\usepackage{xcolor}
\usepackage{tikz}
\definecolor{darkGreen}{rgb}{0,0.6,0}

\def \be{\begin{equation}}
\def \ee{\end{equation}}
\def \ba{\begin{array}}
\def \ea{\end{array}}
\def \bea{\begin{eqnarray}}
\def \eea{\end{eqnarray}}

\newcommand{\Dmun}{{\ensuremath{\Delta \mu^\mathrm{s}_n}}}

\begin{document}


\title{Reexamining the renormalization group: Period doubling onset of chaos}

\author{Archishman Raju}
\author{James P Sethna}
\email{sethna@lassp.cornell.edu}
 \affiliation{Laboratory of Atomic and Solid State Physics, Cornell University, Ithaca, New York 14853-2501, USA}

\date{\today}

\begin{abstract}
We explore fundamental questions about the renormalization group through
a detailed re-examination of Feigenbaum's period doubling route to chaos.
In the space of one-humped maps, the renormalization group characterizes
the behavior near any critical point by the behavior near the fixed point.
We show that this fixed point is far from unique, and characterize a
submanifold of fixed points of alternative RG transformations. We build
on this framework to systematically distinguish and analyze the allowed
singular and `gauge' (analytic and redundant) corrections to scaling,
explaining numerical results from the 
literature. Our analysis inspires several conjectures for critical
phenomena in statistical mechanics.
\end{abstract}

\maketitle
\section{Introduction}


The renormalization group (RG) describes continuous phase transitions exhibiting
scale-invariant fluctuations by a flow in a {\em system space} under a
transformation that coarse-grains and rescales. Systems at their critical
point form a {\em critical manifold} in system space. Points on the
critical manifold share universal exponents and scaling functions because
they flow to a common {\em RG fixed point}.

We shall address three fundamental questions here, each with a long
history in the renormalization group.
\begin{itemize}
\item[\#1] Is the RG fixed point unique? If not, 
\item[\#2] Can any critical point serve as an RG fixed point? If not, 
\item[\#3] What differentiates the subset of critical points
that can serve as RG fixed points?
\end{itemize}
We shall answer these questions in the
context of the period-doubling onset of chaos. We shall then apply these
answers to characterize the corrections to scaling for systems that are
near to the critical point. Our examination of period doubling is inspired by parallel questions in thermodynamic critical phenomena. The results of the analysis of period doubling inspires some conjectures for thermodynamics.

\section{Feigenbaum's RG for period doubling}

\begin{figure}[ht]
\begin {center}
		\includegraphics[width = 0.99\linewidth]{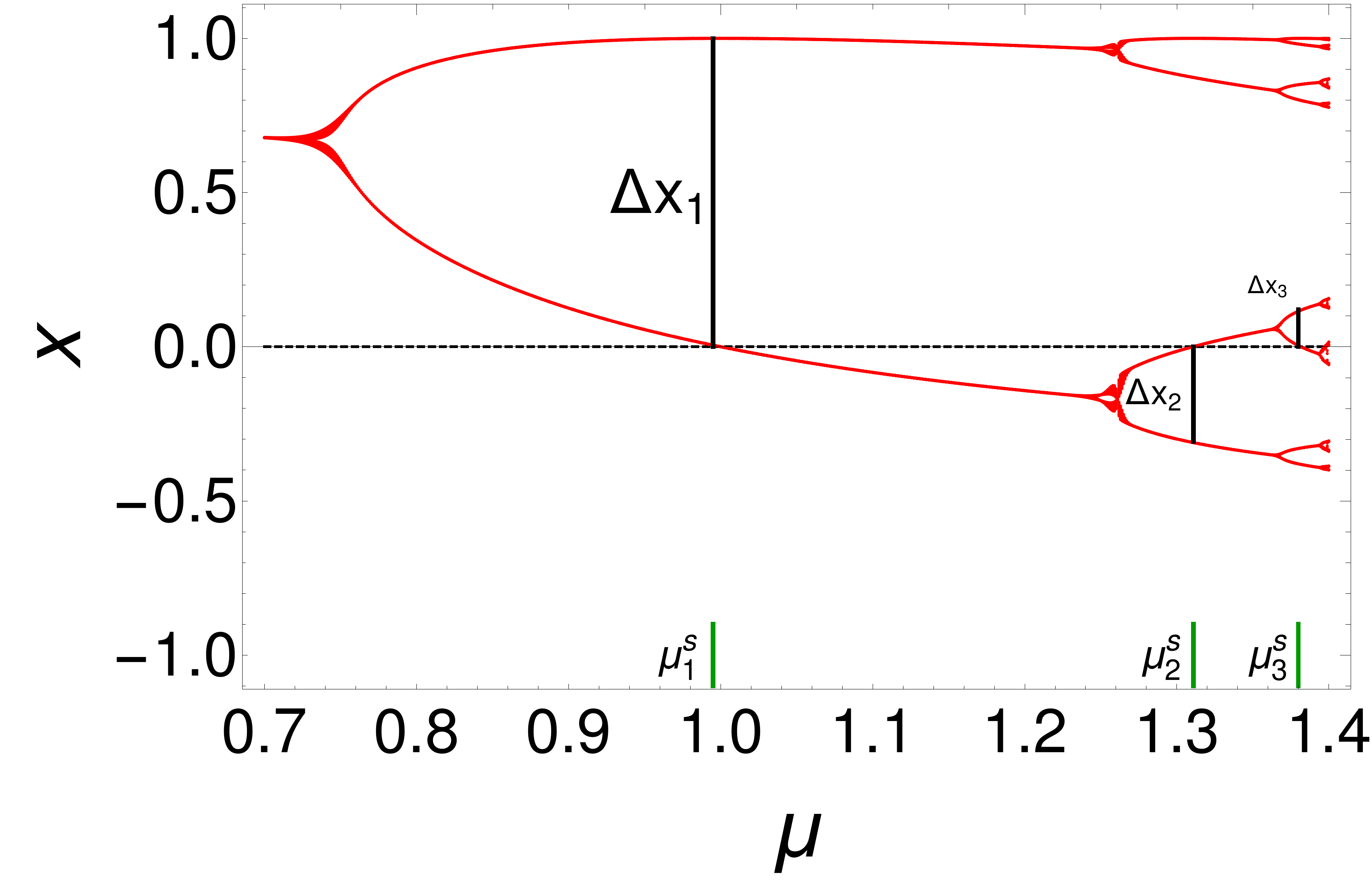}
	\caption{A figure of the bifurcation diagram of the period doubling transition. This is a plot of the fixed points and $n$ cycles at different values of $x$ as a function of the parameter $\mu$ of the map given in the text. The intersection of the bifurcation diagram with the dashed line at $x = 0$ gives the super-stable orbits. There are two predictions in the RG, the spacing between subsequent superstable points given by $\Dmun$ and the vertical distance between the $2^{n - 1}$ cycle and the line $x = 0$ given by $\Delta x_n$ for which we derive the scaling form. The vertical width of the lines in blue give the $\Delta x_n$ and the horizontal coordinate of the green markers gives the values of $\mu_n^{\mathrm{s}}$ (see main text for definitions). }
\label{perioddoubling}
\end {center}
\end{figure}

The period doubling transition is a famous example of the application of the RG to dynamical systems theory. The form of the RG was first worked out by Feigenbaum~\cite{feigenbaum1978quantitative} who showed how the behavior of a a class of iterated maps had universal characteristics. Since then, this kind of analysis has been extended and applied to other maps~\cite{bensimon1986renormalization, rand1982universal}. The archetypal example is that of the logistic map defined by $f(x) = 4 \mu x (1 - x)$. It is conventional to translate the map so that the maximum is at the origin rather than at $x = 1/2$. The symmetrized map that we use in this paper is then defined by  
\begin{equation}
 g(x) = 1 - \mu x^2
\end{equation}
For small values of $\mu$, there is a stable fixed point at some value $x^*$. As the value of the parameter $\mu$ is increased, a stable 2-cycle appears followed by a sequence of bifurcations at parameter value $\mu_n$ with a stable $2^n$ cycle. Eventually, this sequence converges to a point $\mu_\infty$ where you get a transition to chaotic motion. The bifurcation diagram for the period doubling transition is showing in the Figure~\ref{perioddoubling}.  The sequence of $\mu_n$ converge geometrically, $\mu_n - \mu_\infty = \Delta \mu_n \propto \delta^{-n}$, with $\delta$ being a universal constant. It is easier to calculate the scaling form of the superstable points $\mu_n^\mathrm{s}$ which occur roughly midway between two bifurcation points. It is also possible to consider the scaling of the `results' variable $x$, the distance between the $2^{n -1}$ cycle and the line $x = 0$ has a leading order behavior $\Delta x_n \propto 1/\alpha^n$ (see Figure~\ref{perioddoubling}). The bifurcation diagram displays both  universality and self-similarity. Any other one-humped map with a quadratic maximum shows the same sequence of bifurcations. 

The RG operator in period doubling for the `translated' map is given by both coarse graining and rescaling, $\mathcal{T}[g(x)] = \alpha g(g(x/\alpha))$. 
The fixed point is given by the whole function $g^*(x)$. The operator $\mathcal{T}$ can be linearized close to the fixed point. Its largest eigenvalue $\delta$ explains the leading scaling behavior $\Dmun \propto 1/\delta^n$. Here, we will consider corrections to this result with an aim to answer some of the questions raised in the first paragraph. 

\section{Is the fixed point unique?}

First, let us give the answer in thermodynamics. There are several different ways to coarse grain a physical system.
Coarse graining in momentum space leaves you at a fixed point which is
different from a fixed point that coarse grains in real space. For example,
anisotropy due to the lattice for an Ising model will not vanish in the
real space
renormalization group (RG) whereas it will be washed out if the coarse
graining is done with spherical cutoffs in momentum space. Thus, the
momentum space and real space RG must lead to different fixed point
Hamiltonians. This answers question~\#1: The fixed point is not unique.  

The fixed point can be moved by changing variables in the degrees of freedom ({\em i.e.}, not the control parameters). Such a change of variables leads to `redundant variables' explored by Wegner~\cite{wegner1974some} in detail. They arise when considering a change of coordinates in the degree of freedom. So, for example, the cubic term in the Hamiltonian of the Ising model does not contribute to the scaling behavior because it can be removed by a change of coordinates~\cite{cardy1996scaling}. The statistical mechanics literature therefore usually ignores the effect of these variables on the scaling behavior.

What is the equivalent of these redundant variables in period doubling? Let a change of coordinate be $y = \phi(x)$. This induces a map $\tilde{g}(y) = \phi(g(\phi^{-1}(y)))$. Naturally, this leads to a new fixed point function $\tilde{g}^*(y) = \phi(g^*(\phi^{-1}(y)))$. The space of transformations $\phi$ thus generates a family of fixed points. Each of these fixed points have an associated renormalization group flow $\tilde{T} = \phi \circ T \circ \phi^{-1}$. Since the fixed points depend on a coordinate choice $x$, we choose to call the associated corrections \textit{gauge corrections to scaling}.

Gauge invariance in electromagnetism can be viewed as a spatially-varying change of coordinates in the phase of the quantum wave function. Gauge transformations in general relativity are just coordinate transformations~\cite{FlanaganH05}. Coordinate changes in other systems (e.g., models of moving interfaces~\cite{LangerGJ92}) are also naturally described as gauge transformations. Here the universal predictions correspond to those quantities that are gauge-invariant -- independent of changes of coordinates in the parameters (either control parameters or predictions of the theory). Those corrections to scaling that depend on the choice of coordinates ({\em e.g.}, not gauge invariant) we shall call {\em gauge corrections to scaling}, to distinguish them from {\em singular corrections to scaling} due to fundamentally new irrelevant operators under the RG, and from logarithmic and other anomalous scaling forms due to universal but nonlinear terms in the renormalization-group flow~\cite{raju2017renormalization}.


\section{Can any critical point serve as an RG fixed point?}
 This leads to question~\#2: Can any critical point serve as an RG fixed point? In thermodynamics, an early numerical study by Swendsen was based on
changing the form of the RG to change the 2D Ising critical point to be the
fixed point~\cite{swendsen1984optimization}. However, Fisher and Randeria~\cite{fisher1986location} argued that the fixed point was distinguished, up to redundant variables, as the point that has no singular corrections to scaling. The singular corrections to scaling came from irrelevant variables with universal critical exponents.%
  \footnote{We shall conjecture later that all of the irrelevant corrections
  to scaling for the 2D nearest neighbor Ising model are removable by 
  changing the form of the renormalization group; Swendsen's assumption
  was not generally true, but may have been true for the model he studied.}
 
 To answer this question in period doubling, we consider an arbitrary change of coordinates $y = \phi(x)$ and apply the original renormalization group transformation $\mathcal{T}$  on the new fixed point $\tilde{g}^*$ (following Ref.~\cite{feigenbaum1979universal}), generating a space of `redundant' (gauge) variables. 
\begin{equation}
 \mathcal{T}[\tilde{g}^*](x) = \alpha(\phi(g^*(g^*(\phi^{-1}(\alpha^{-1} x)))))
 \label{periodeq}
\end{equation}

To make progress, let the coordinate change have the infinitesimal form $\phi(x) = x + \epsilon \Psi(x)$. The inverse transformation $\phi^{-1}(x) = x - \epsilon \Psi(x)$. At the fixed point $\alpha (g^*(g^*(x/\alpha))) = g^*(x)$. Taking a derivative of this equation gives ${g^*}'({g^*}(x/\alpha))){g^*}'(x/\alpha) = {g^*}'(x)$. We can expand to linear order in $\epsilon$
\begin{align}
 \mathcal{T}[\tilde{g}^*](x) &= \alpha(\phi(g^*(g^*(\phi^{-1}(x/\alpha))))) , \\
 &= g^*(x) + \alpha \epsilon \left( \Psi (g^*(x)/\alpha) - \left(\Psi(x/\alpha) \right) {g^*}'(x) \right) .
\end{align}

Let $\Psi(x)$ have a Taylor series $\Psi(x) = \sum_p \Psi_p x^p$. We then get 
\begin{equation}
 \mathcal{T}[\tilde{g}^*](x) - g^*(x) = \sum_p \alpha^{1 - p} ({g^*(x)}^p - {g^*}'(x) x^p) .
\end{equation}


Hence the space of such equivalent fixed points have eigenvalues $\alpha^{1 - p}$ with eigenfunctions given by $g^*(x)^p - {g^*}'(x) x^p$. The odd eigenvalues%
  \footnote{The period doubling literature often restricts itself to an even subspace of functions which does not see half of the eigenvalues. This is why we are reporting only half of the eigenvalues here.}
are given numerically by $\{1.00, 0.16, 0.0255, ...\}$. Feigenbaum, in
his original derivation of these irrelevant
eigenfunctions~\cite{feigenbaum1979universal}, conjectured
that they spanned the function
space -- that all of the irrelevant
eigenvalues in period doubling are gauge-irrelevant in our nomenclature
(a conjecture equivalent to Swendsen's assumption). 
A numerical computation of the eigenvalues at the fixed point,
$\{4.67, \underline{1.00}, \underline{0.160}, -0.124, -0.0573, \underline{0.0255},...\}$~\cite{christiansen1990spectrum}, shows that only some of them
are given by by powers of $\alpha$ (`gauge' eigenvalues underlined above). Others are fundamentally new numbers which cannot be written down in terms of already known eigenvalues (singular irrelevant eigenvalues). The only relevant eigenvalue is the first one given by $\delta \approx 4.67$. 
Thus not all critical points can serve as RG fixed points.


Our analysis of period doubling suggests an alternative approach to the understanding of redundant variables in statistical mechanics. Redundant variables are (usually) irrelevant variables which contribute to the corrections to scaling of the results of the RG. However, they do not lead to any new eigenvalues. Their `gauge' eigenvalues can be written in terms of some linear combination (or in this discrete case, by some product) of already known eigenvalues. Other irrelevant variables that have fundamentally new eigenvalues contribute genuine \textit{singular} corrections to scaling. Hence, some of the variables in the renormalization group are like a gauge choice. Having fixed a gauge, it contributes to the observed behavior. Thus we discriminate between genuine \textit{singular} corrections to scaling and \textit{gauge irrelevant} corrections to scaling, both of which come from irrelevant variables.%
  \footnote{In fact the cubic term in the Hamiltonian of the Ising model is usually quoted to have an eigenvalues $d - \lambda_h$ where $\lambda_h$ is the eigenvalue of the linear term, consistent with our classification.}
Based on this analysis, we conjecture that Randeria and Fisher's answer to question~\#3 is general: Critical points
that can serve as RG fixed points have no singular corrections to scaling, 
excluding gauge-irrelevant corrections due to redundant coordinate
changes in the results.
\begin{figure*}[ht]
\captionsetup[subfigure]{justification=centering}
  \begin{subfigure}[b]{.3\linewidth}
\centering
    \includegraphics[width=.8\textwidth]{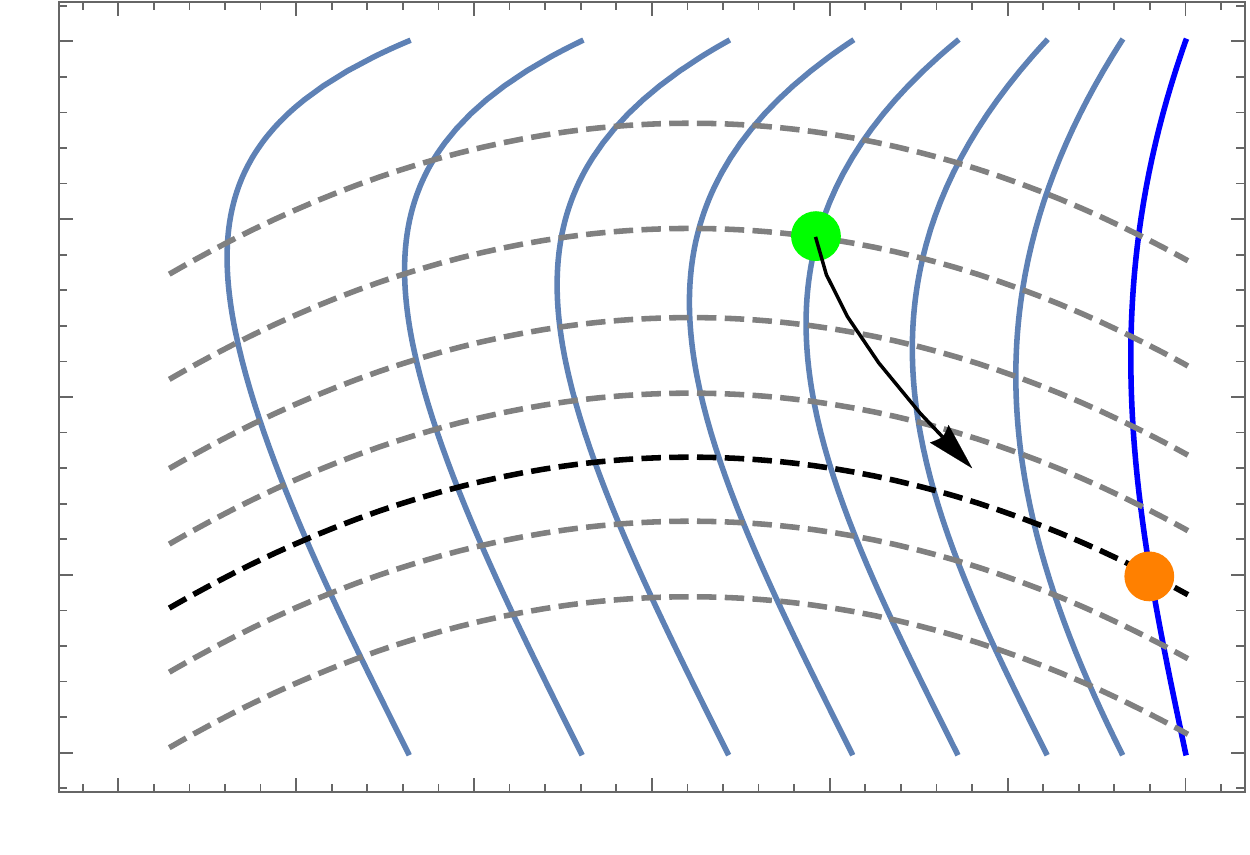}
    \caption{}\label{gauge1}
  \end{subfigure}%
  \begin{subfigure}[b]{.3\linewidth}
  
    \includegraphics[width=.8\textwidth]{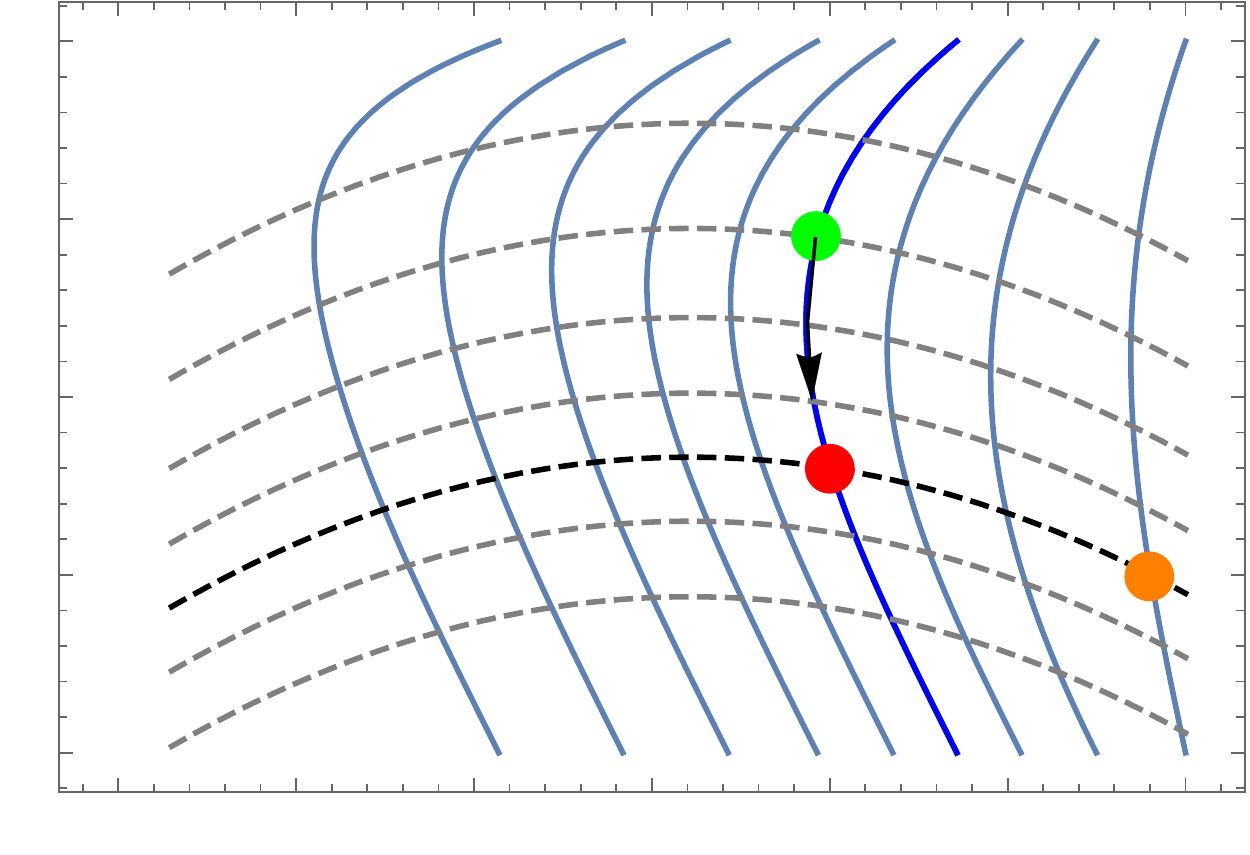}
    \caption{}\label{gauge2}

  \end{subfigure}
    \begin{subfigure}[b]{0.3\linewidth}
    
    \includegraphics[width=.8\textwidth]{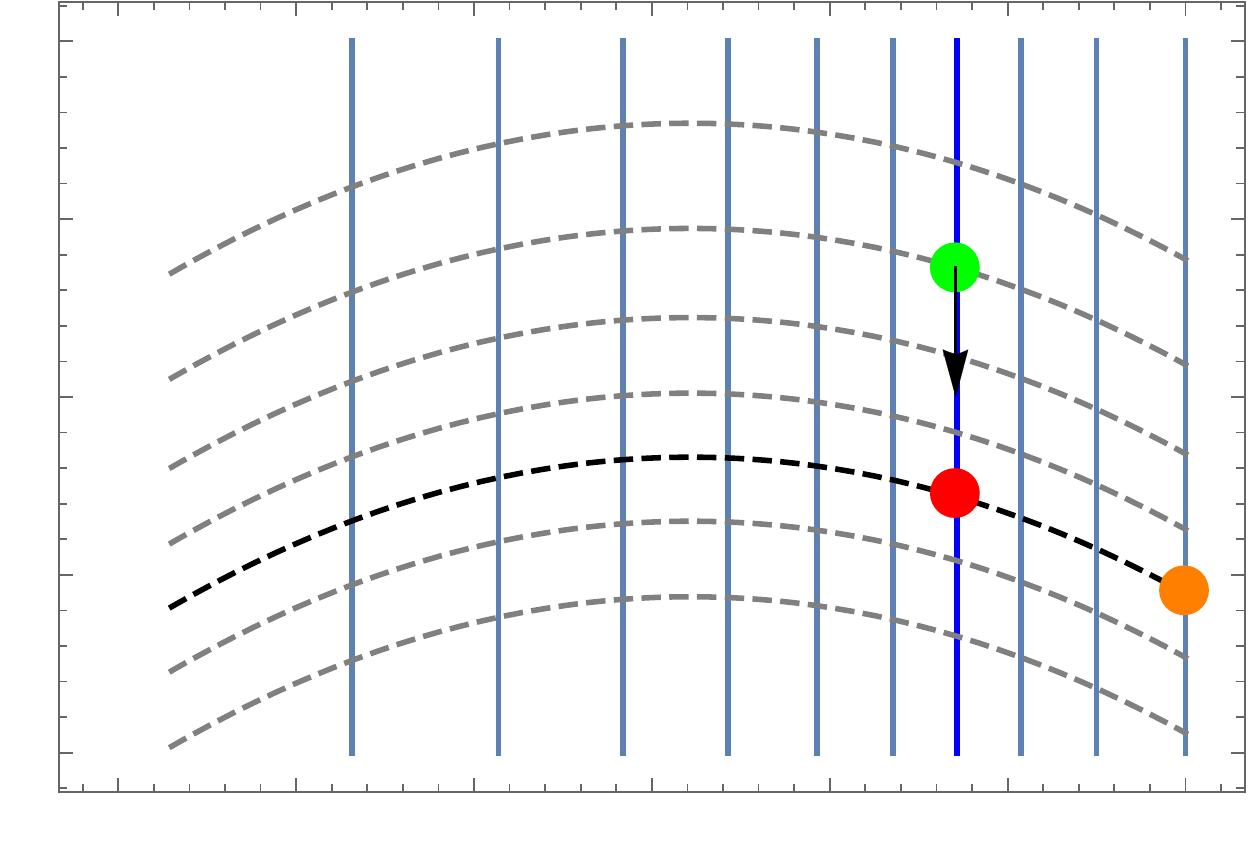}
    \caption{}\label{gauge3}
  \end{subfigure}\\%
\caption{{\bf Renormalization group on the critical manifold: changing coordinates.}
(a)~The traditional RG predicts scaling behavior for the logistic map (green dot) because it flows geometrically to a fixed point (orange dot) under rescaling. The behavior is universal because almost all critical one-humped maps flow to the same fixed point.  The {\bf dark} solid line is a singular irrelevant direction without gauge corrections.  The {\em dark} dashed line corresponds to gauge (redundant) perturbations; it is the submanifold of allowed fixed points for different RG choices. 
(b)~By changing coordinates in the results variable $y=\phi(x)$ we can flow to a different fixed point (red dot). For the coordinate choice shown,
the gauge corrections to scaling vanish for the logistic map (green dot).
The light solid lines, flowing along singular directions to the various
RG fixed points, foliate the space into submanifolds. 
(c)~Finally, we can change to {\em normal form} coordinates 
(Section~\ref{normalformsec}). Here the flow at the RG fixed point is hyperbolic, 
and the normal flow on the singular submanifold becomes completely linear.}
\label{gaugefigures}
\end{figure*}
 




\section{Singular and redundant foliations of critical manifold: Normal
form theory.}
\label{normalformsec}

 Our work here is part of a larger effort to systematize corrections to scaling in critical phenomena by using normal form theory~\cite{raju2017renormalization}. In previous work, we considered changing coordinates in the control parameters. These changes of coordinates systematically generate the singularity at the critical point (including singular corrections) and the analytic and redundant
corrections to scaling (which we call `gauge' corrections).

Corrections to scaling in statistical mechanics were first explained by Wegner~\cite{Wegner72} and Aharony and Fisher~\cite{aharony1983nonlinear} who physically interpreted nonlinear terms in the RG as analytic corrections to scaling. Analytic corrections to scaling come from nonlinear terms in the RG or from coordinate changes in the control variables (like the magnetic field and temperature in an Ising model, and $\mu$ in period doubling). Normal form theory allows us to see the equivalence of these two. The redundant variables, mentioned above, come from coordinate changes in the results (like magnetization in the Ising model,
or $x$ in period doubling). They correspond to eigendirections which are (usually) irrelevant. Analytic and redundant variables are treated completely differently by the RG, but physically have very similar origins. The model and experiment use different coordinate systems to control and measure physical quantities.  

The period doubling case allows one to not only see this similarity but also to discriminate between redundant (gauge) and singular corrections
(see Figure~\ref{gaugefigures}). A generic system has complicated RG flows with its fixed point having both singular and gauge corrections to scaling. We first choose an appropriate set of coordinates in the results variable that moves the fixed point to set all of the gauge corrections to zero. This confines the flow to a manifold which has only singular corrections. The set of all such manifolds which correspond to different fixed points foliate the complete space. In this manifold where all gauge variables are set to zero, we change coordinates in the parameters to the normal form that make the flows as simple as possible~\cite{raju2017renormalization}. This gives the leading singularity at the critical point which experimental data can be fit to. Finally, corrections to scaling can be systematically incorporated by letting the normal form coordinates be an arbitrary Taylor series of the experimental coordinates.  
 
 If the predictions that we are making involve the results variable $x$, then the experimental data will generically have all possible corrections to scaling (including all of the gauge corrections). If, however, the predictions only involve the parameter $\mu$, then the gauge corrections in the results will not matter. We now illustrate the above procedure by deriving the full corrections to scaling for the RG of period doubling.

\section{Corrections to scaling in period doubling:
	 singular {\em vs.} gauge.}

 Corrections to scaling in period doubling have been considered before~\cite{mao1987corrections, briggs1994corrections, damgaard1988non,reick1992universal}. While an ad-hoc form of the corrections was presented in Ref.~\cite{mao1987corrections}, the singular corrections to scaling coming from the irrelevant eigenvalues within the linear RG was derived in  Refs.~\cite{briggs1994corrections,reick1992universal}. Here, we derive a more complete form of the corrections to scaling to compare how singular and gauge corrections appear in the physical predictions. As explained above, we set the gauge corrections to zero by choosing coordinates appropriately. Then, on the manifold with no gauge corrections to scaling, we move to \textit{normal form} coordinates which linearize the RG flow.

Going to the normal form coordinates leads to an enormous simplification. The RG has nonlinear terms which are now absorbed in a coordinate change. We explain when this can be done in Ref.~\cite{raju2017renormalization}. Briefly, we can absorb all nonlinear terms in the RG in the absence of resonances. These resonances happen for continuous (discrete) RG flows when certain integer combinations (products) of eigenvalues are zero (one). We hypothesize that the singular corrections to scaling in period doubling have no resonances. In this case, the flow can be completely linearized and the fixed point is called hyperbolic. We use the existence of a coordinate system where the flow is linear to characterize the corrections to scaling.


Let us start by considering corrections to scaling for the values $\Dmun$
in Fig.~\ref{perioddoubling}.
We denote the linearization of $\mathcal{T}$ by $\mathcal{T}_L$. The critical point is at the value of $\mu = \mu_\infty$.  Let $\Delta \mu = \mu - \mu_\infty$. We denote the normal form coordinates with a $\sim$. We denoted the redundant eigenfunctions above by $\Psi_p(x)$. We will denote the eigenfunctions which are genuinely singular by $\Phi_p(x)$ and the associated eigenvalues by $\lambda_p$. In our coordinates, then  
\begin{equation}
\label{eq:NonRedundantDeltaMuRG}
 \mathcal{T}_L[g_\mu - g^*](x) = \Delta \mu  \sum_p \tilde{a}_p \lambda_p \Phi_p (x) 
\end{equation}
Now, let us act with the operator $n$ times, so 
\begin{equation}
 \mathcal{T}_L^n[g_\mu - g^*](x) = \Delta \mu \sum_p \tilde{a}_p \lambda_p^n \Phi_p (x) 
\end{equation}
If $g$ has a $2^n$ cycle, with $\mu = \mu_n$, then the application of $\mathcal{T}^n$ has a defined value at $x = 0$, so 
\begin{equation}
 \sum_p \Dmun \tilde{a}_p \lambda_p^n \Phi_p(0) = c ,
\end{equation}
where $c$ is a constant. 
We redefine constants $\tilde{a}_p \rightarrow \tilde{a}_p/\Phi_p(0)$ to absorb $\Phi_p(0)$. 
This gives
\begin{equation}
 \sum_{p} (\tilde{a}_p \Dmun) \lambda_p^n = 1 . 
\end{equation}

\begin{figure}[ht]
\begin {center}
		\includegraphics[width = 0.99\linewidth]{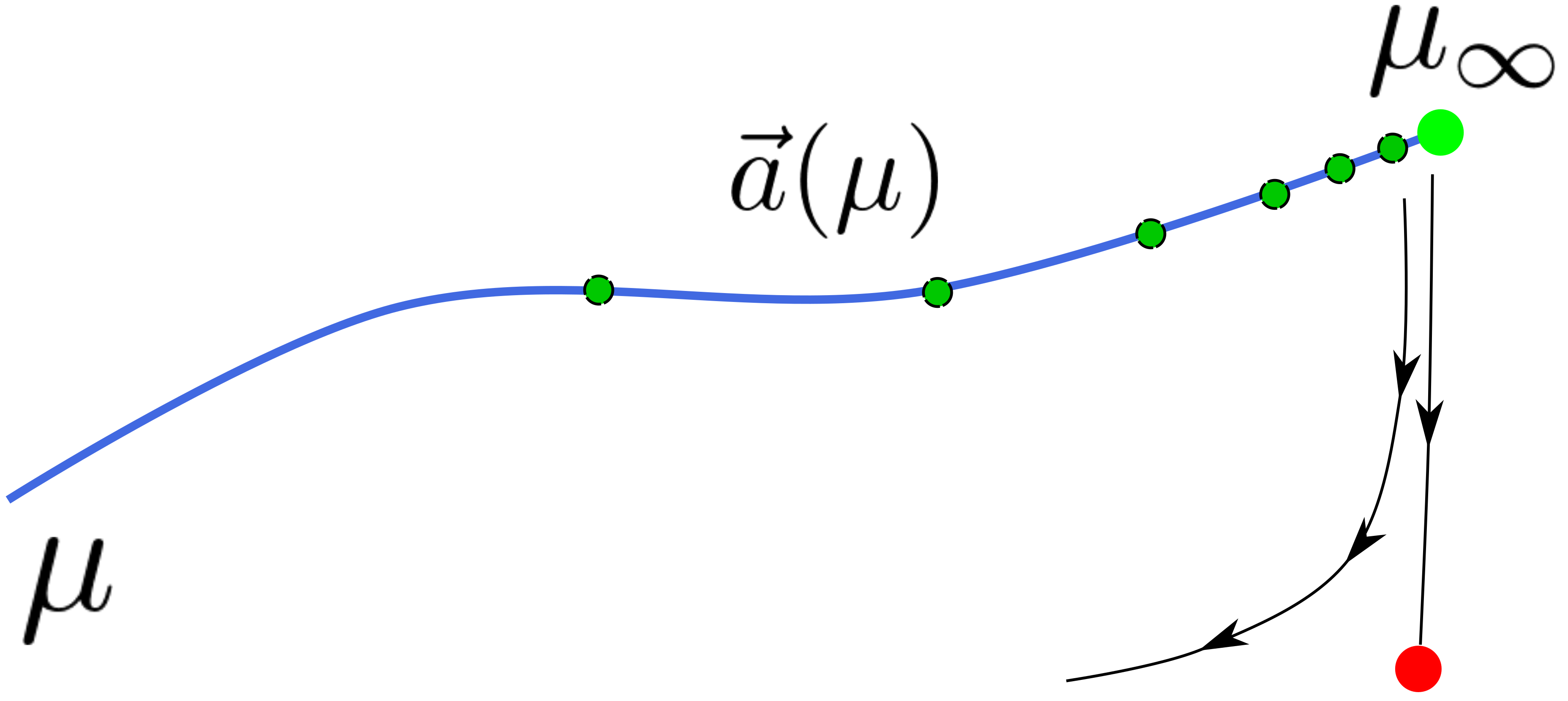}
	\caption{The blue line shows the function parameterized by $\mu$ with a fixed point at $\mu_\infty$ shown as a green disk. The $n$ cycles are shown as smaller dark green disks lie geometrically spaced on the line. The vector $\vec{a}(\mu)$ gives the amplitude of the irrelevant eigenfunctions as a function of $\mu$. The red disk is the fixed point in the space with redundant variables set to 0. The RG flows are shown in black.}
\label{perioddoublingflow}
\end {center}
\end{figure}

Now, to include any corrections from the nonlinear regime of the RG, we assume that the coefficients $\tilde{a}_p$ 
have a series expansion $\tilde{a}_p = a_p^{(0)} + a^{(1)}_p \Dmun + ...$. This is pictorially represented in Figure~\ref{perioddoublingflow}. Along the line parameterized by $\mu$ the function $g_\mu (x)$ has different amplitudes of the coefficients $a_p (\mu)$. So, the final expression now implicitly gives the scaling behavior of $\Dmun$
\begin{equation}
 \Dmun = \frac{1}{(\sum_{p} (a^{(0)}_p + a^{(1)}_p \Dmun + ...) \lambda_p^n)} .
 \label{correctionfullexpression}
\end{equation}
 The above equation can be solved order by order in $\Delta \mu^s_n$. The most useful result is directed at an experimentalist, what are the terms that give the corrections to scaling and how many independent coefficients are there to fit to the results? The expression in Equation~\ref{correctionfullexpression} is best studied perturbatively. The lowest order expression is 
\begin{equation}
 \Dmun = \frac{1}{(a^{(0)}_1 \delta^n)} .
\end{equation}
We use the freedom to rescale $\Delta \mu$ to set $a^{(0)}_1 = 1$. We will organize terms by powers of $\delta^{-n}$. Then the next order expression is 
\begin{equation}
 \Dmun = \frac{1}{\delta^n} \left(1 - \frac{a^{(1)}_1}{\delta^n} + \sum_{p \neq 1} a_p^{(0)} \lambda_p^n/\delta^n \right) . 
\end{equation}
So far, the number of independent coefficients equal the number of new terms. At the next order, we get a much more complicated expression. The expression for $\Dmun$ is 
\begin{align}
 \Dmun &= \frac{1}{\delta^n} \left (1 - \frac{1}{\delta^n} \left( \sum_{p \neq 1} \lambda_p^n a_p^{(0)} + a_1^{(1)} \right)\right) \nonumber \\ &+ \frac{1}{\delta^{3 n}} \left (\sum_{p, q \neq 1} \lambda_p^{n} \lambda_q^{n} a_q^{(0)} a_p^{(0)} \right) \nonumber \\ &+ \frac{1}{\delta^{3 n}} \left (3 \sum_{p \neq 1} \lambda_p^n (a^{(0)}_p a_1^{(1)} - a^{(1)}_p) + (2 ({a_1^{(1)}})^2 - a_0^{(2)}) \right) 
\end{align}
As can be seen, corrections to this order add more terms than coefficients (e.g. if we kept only two of the irrelevant eigenvalues, it would lead to 6 new terms but only three new coefficients). The coefficients of the various terms have a somewhat complicated relationship between them. We expect this to be true at higher orders as well though we have not yet found a general expression for the corrections to scaling at arbitrary order. At each order, the corrections to scaling with the relationship between the various terms can be derived perturbatively. There are several things to notice in this expression for $\Dmun$. One, there are certain terms that go as $\frac{1}{\delta^{i n}}$ for integer $i$. These can be viewed as analytic corrections to the relevant variable $\mu$. Second, the correction to scaling does not involve the result in any way and so it should be independent of the gauge corrections to scaling in the results. Changing coordinates in $x$ should not affect the expression for $\Dmun$. Hence, powers of $\alpha$ should not be observed in the corrections to scaling for $\Dmun$ (that is, Equation~\ref{eq:NonRedundantDeltaMuRG} does not include the redundant eigendirections $\Psi_p$).
The corrections due to the eigenvalue $\alpha^0$ coincides with analytic corrections to scaling to $\mu$ and those analytic corrections can still appear -- analytic corrections (usually analyzed as changes of coordinates in the control parameters) are here also gauge corrections to scaling. 
Ref.~\cite{briggs1994corrections} observed this result numerically to lowest
order for the logistic map. 


We can similarly derive a form for the corrections to the scaling of $g^{2^{n - 1}}(0)$ which we call $\Delta x_n$ following Ref.~\cite{mao1987corrections} (see Figure~\ref{perioddoubling}). Asymptotically, these are just given by $\alpha^{-n}$. To derive the corrections, we notice that $\alpha^n g^{2^{n - 1}} (\alpha^{1-n} x)$ is the same as acting the operator $\mathcal{T}$ and so has a similar expansion 
\begin{equation}
 \alpha^n g^{2^{n - 1}} (\alpha^{1-n} x) = g^*(x) + \Dmun \sum_p a_p \lambda_p^{n-1} \Phi_p (x) .
 \end{equation}
 Evaluating this at $x = 0$ gives 
 \begin{equation}
  \Delta x_n = \alpha^{-n} (1 +  \Dmun \sum_p \tilde{a}_p \lambda_p^{n-1} \Phi_p (0)) .
 \end{equation}
  We can include the corrections to scaling from the gauge variables by assuming that the gauge and singular directions can simultaneously be brought to a normal form (this would give a rectangular grid instead of Fig.~\ref{gauge3}). There is a subtlety here: gauge variables can not lead to any new singularities and hence do not have any resonance terms in their flow even if their RG eigenvalues have resonances~\cite{raju2017renormalization}. Including the gauge  corrections to scaling then gives
  \begin{align}
  \Delta x_n &= \alpha^{-n} (1 +  \Dmun \sum_p \tilde{a}_p \lambda_p^{n-1} \Phi_p (0) \nonumber \\ &+ \Dmun \sum_p \tilde{b}_p \alpha^{(1- p) (n-1)} \Psi_p(0)) . 
  \end{align}

Substituting the form of $\Dmun$ and using the series of expansion of $a_p$ gives the full form of the corrections to scaling of $\Delta x_n$. In this case, the gauge corrections to scaling in the results \textit{should} affect the value of $\Delta x_n$ and contribute to the observed behavior as is indeed seen to lowest order in Ref.~\cite{briggs1994corrections}. In Ref.~\cite{damgaard1988non}, a change in scaling behavior of $\Delta x_n$ was seen under a change of coordinates though a RG explanation was not given. The interpretation becomes clear here, a change of coordinates will affect the gauge corrections to scaling of $\Delta x_n$ and hence change its scaling behavior. 

 Since the leading scaling behavior of $\Delta x_n$ is $\alpha^{-n}$ and all of the gauge eigenvalues are $\alpha^{1 - p}$ for integer $p$, the gauge corrections can simply be generated as analytic corrections in $\Delta x_n$.  This has a parallel in the 2-d Ising model where the corrections to scaling coming from irrelevant variables predicted by conformal field theory cannot be distinguished from analytic corrections. All of the conformal field theory predictions are for `descendant operators' which are obtained by taking derivatives of primary (relevant) operators. These operators have integer eigenvalues. Normal form theory suggests that they generically should lead to logarithmic powers which are not observed in the square lattice 2-d Ising model. Meanwhile, the leading genuine singular correction to scaling coming from an operator with eigenvalues $-4/3$ seems to have zero amplitude in the 2-d square lattice Ising model. Thus, Barma and Fisher~\cite{barma1984corrections,
barma1985two} had to use a double-Gaussian model to find evidence for a genuine singular correction. Our analysis here would suggest that the irrelevant operators predicted by conformal field theory, and observed in the 2-d Ising model are all contributing gauge corrections to scaling (Swendsen was correct for his
model), whereas the irrelevant variable with eigenvalue $-4/3$ that Barma and Fisher observe is a genuine singular correction to scaling.

\section{Conclusion}

We have examined some deep questions about the renormalization group in the context of period doubling. We showed that there is some freedom to move the fixed point of the RG associated with gauge transformations in the coordinates of the map. We have also  derived the full form of the corrections to scaling of the period doubling transition. In doing so, we propose a strategy for systematically predicting corrections to scaling at critical points. One first goes to the sub-manifold with no gauge correction to scaling and then to normal form coordinates. Predictions of the RG which do not involve the traditional `results' variables are unaffected by the gauge corrections. For the results however, the gauge corrections do contribute in a manner similar to but yet distinct from other universal singular corrections to scaling. Our explicit analysis of the RG in period doubling allows us to explicitly distinguish between genuine singular corrections to scaling and gauge corrections which can be removed by coordinate changes. Rather than changing coordinates to get rid of the gauge corrections, they can be retained in the analysis and are distinguished by the fact that they lead to no new universal eigenvalues but still contribute to the corrections to scaling in the results. 

We conjecture that the difference between such gauge eigenvalues, and the universal eigenvalues associated with the RG lies in the fact that these gauge eigenvalues are some combination of already known eigenvalues of the RG. The 2-d Ising model is an interesting example where this conjecture can be tested. In period doubling, the degree of freedom $x$ and the parameter $\mu$ are both one dimensional and corrections to scaling coming from changing variables in either of them are easily derived.

In statistical mechanics, corrections to scaling due to coordinate changes in 
the control variables ({\em e.g.}, temperature and field in the Ising model) are
usually termed {\em analytic} corrections to scaling~\cite{aharony1983nonlinear}. Corrections to scaling involving changing the results variables
({\em e.g.}, the definition of spin or magnetization) are traditionally
 termed {\em redundant}
corrections to scaling. Here, noting the close analogy between these corrections
we propose to denote both types of corrections as {\em gauge} corrections
to scaling (changing as we measure, or gauge, the various fields differently). 
We also note that many irrelevant variables in the renormalization group
are indeed due to gauge degrees of freedom in the results variables, and
conjecture that these gauge-irrelevant eigenvalues and corrections to scaling
are best be understood as combinations of relevant and singular-irrelevant
eigenvalues. 

\begin{acknowledgments}
This work was supported by an NSF grant DMR-1719490. 
\end{acknowledgments}


\bibliographystyle{unsrt}
\bibliography{perioddoublingbib}

\end{document}